\documentclass{moriond}

\usepackage{amssymb}

\def\be{\begin{equation}}
\def\ee{\end{equation}}
\def\bea{\begin{eqnarray}}
\def\eea{\end{eqnarray}}

\newcommand{\Photo}{}

\begin{document}
\vspace*{4cm}
\title{Short Baseline Oscillations and the Gallium Mystery}

\author{Vedran Brdar}

\address{Theoretical Physics Department, CERN,\\
Esplande des Particules, 1211 Geneva 23, Switzerland}

\maketitle\abstracts{
Data from several neutrino experiments suggest an anomalous neutrino flavor transition across relatively short baselines which is in conflict with the three-flavor neutrino oscillation paradigm. In particular, MiniBooNE and BEST collaborations have reported anomalous findings at $\sim 5\sigma$. In this contribution, such measurements and their possible explanations within and beyond the Standard Model are discussed.}

\section{Introduction}
\label{sec:intro}
Neutrino oscillation is a Nobel Prize awarded phenomenon \cite{Super-Kamiokande:1998kpq,SNO:2002tuh} which has by now been measured using several different neutrino sources and detection techniques \cite{Esteban:2020cvm}. While this program has firmly established the three-flavor neutrino oscillation paradigm, there are still several experimental hints which suggest that the neutrino sector could possibly be even richer, e.g. supplemented with light sterile neutrino(s) \cite{Giunti:2019aiy}.

In particular, LSND experiment observed  $\sim 3\sigma$ excess of electron antineutrinos from the stopped pion source \cite{LSND:2001aii}. This suggests $\mathcal{O}(10^{-3})$ probability for electron antineutrino appearance from the source of muon antineutrinos. Given the energy of the beam and the baseline,  such measurement can not be explained with the known parameters in the neutrino sector; hence, this motivates beyond the Standard Model (BSM) interpretation of this excess. LSND anomaly was tested at Fermilab with the MiniBooNE experiment which has, on several occasions \cite{MiniBooNE:2008yuf,MiniBooNE:2018esg,MiniBooNE:2020pnu}, also reported a low energy excess in both neutrino and antineutrino channel. In their most recent analysis \cite{MiniBooNE:2020pnu}, $4.8\sigma$ excess is claimed. MiniBooNE anomaly is currently being tested at the Short Baseline Neutrino Program (SBN) \cite{MicroBooNE:2015bmn,Machado:2019oxb} and MicroBooNE collaboration has already released the first results \cite{MicroBooNE:2021zai,MicroBooNE:2021tya}. 

Both LSND and MiniBooNE have recorded accelerator-based neutrino events. The remaining short baseline anomalies are associated to somewhat smaller neutrino energies, namely the MeV scale. The reactor neutrino anomaly \cite{Mention:2011rk,Huber:2011wv} is a reported disagreement between the observed and the expected event rates at detectors placed in the vicinity of nuclear reactors. This effect can be explained via mixing between electron and sterile (anti)neutrino \cite{Dentler:2017tkw,Gariazzo:2018mwd}.
However, including the recent refinements in the reactor flux calculations \cite{Kopeikin:2021ugh} it is strongly suggested that the reactor anomaly is disappearing; for the model with sterile neutrino, now the strong bounds are placed \cite{Giunti:2021kab,Berryman:2021yan}. Regarding reactor antineutrino spectra, it is worthwhile, for completeness, to mention the so called 5 MeV bump \cite{Huber:2016xis} that was reported by several experiments \cite{Seo:2014xei,DayaBay:2015lja,DoubleChooz:2015mfm} and which is very difficult to explain with BSM physics \cite{Berryman:2018jxt}.

 Last but not least, several experiments with $^{71}$Ga as a detection material have observed interactions of neutrinos from very intense radioactive sources and reported a deficit; GALLEX \cite{GALLEX:1997lja} and SAGE \cite{SAGE:1998fvr} produced data that corresponds to not very significant $\sim 2\sigma$. However, very recently,  BEST collaboration performed new measurements \cite{Barinov:2021asz} and the lack of the observed with respect to the expected number of events is now established at a far more significant level. The gallium anomaly stands at $\gtrsim 5\sigma$ \cite{Barinov:2021mjj,Giunti:2022btk,Brdar:2023cms}. 
 
 In Sec.~\ref{sec:MB} and Sec.~\ref{sec:gallium}, the two statistically most significant anomalies among the above discussed, MiniBooNE and Gallium, are examined.

\section{MiniBooNE Anomaly}
\label{sec:MB}
MiniBooNE collaboration reported a $\sim 5\sigma$ excess of electron-like events \cite{MiniBooNE:2020pnu} with reconstructed neutrino energies between $\sim 200$ and $400$ MeV. Let us first introduce the main Standard Model processes that contribute to the appearance of a single electromagnetic shower in the detector. Even though the beam consists of mostly muon neutrinos \footnote{For brevity, in this section, the term neutrino is used for both neutrinos and antineutrinos.}, there is a small admixture of electron neutrinos. Through the charged current interaction, $\nu_e + n \to e^- + p$, electron neutrinos lead to the production of electrons in the detector which in turn induce electromagnetic showers. Another relevant process, which can be realized via scattering of neutrinos of all flavors off nuclei \footnote{ Namely carbon and hydrogen since the detector was filled with mineral oil.}, is the neutral current production of $\pi^0$ which promptly decays to two photons. If the photons are very collimated, or if only one of them converts into $e^+e^-$ pair within the fiducial volume, or if one of the photons is very soft, with energy below  the detection threshold, the signature of the process would be a single electromagnetic shower. Finally, neutrino-nucleus scattering can also lead to a production of hadronic GeV-scale resonances which can decay to photons that appear as electron-like events in the detector.

\begin{figure}
  \centering
  \begin{tabular}{c@{\quad\quad}c}
    \includegraphics[width=0.45\textwidth]{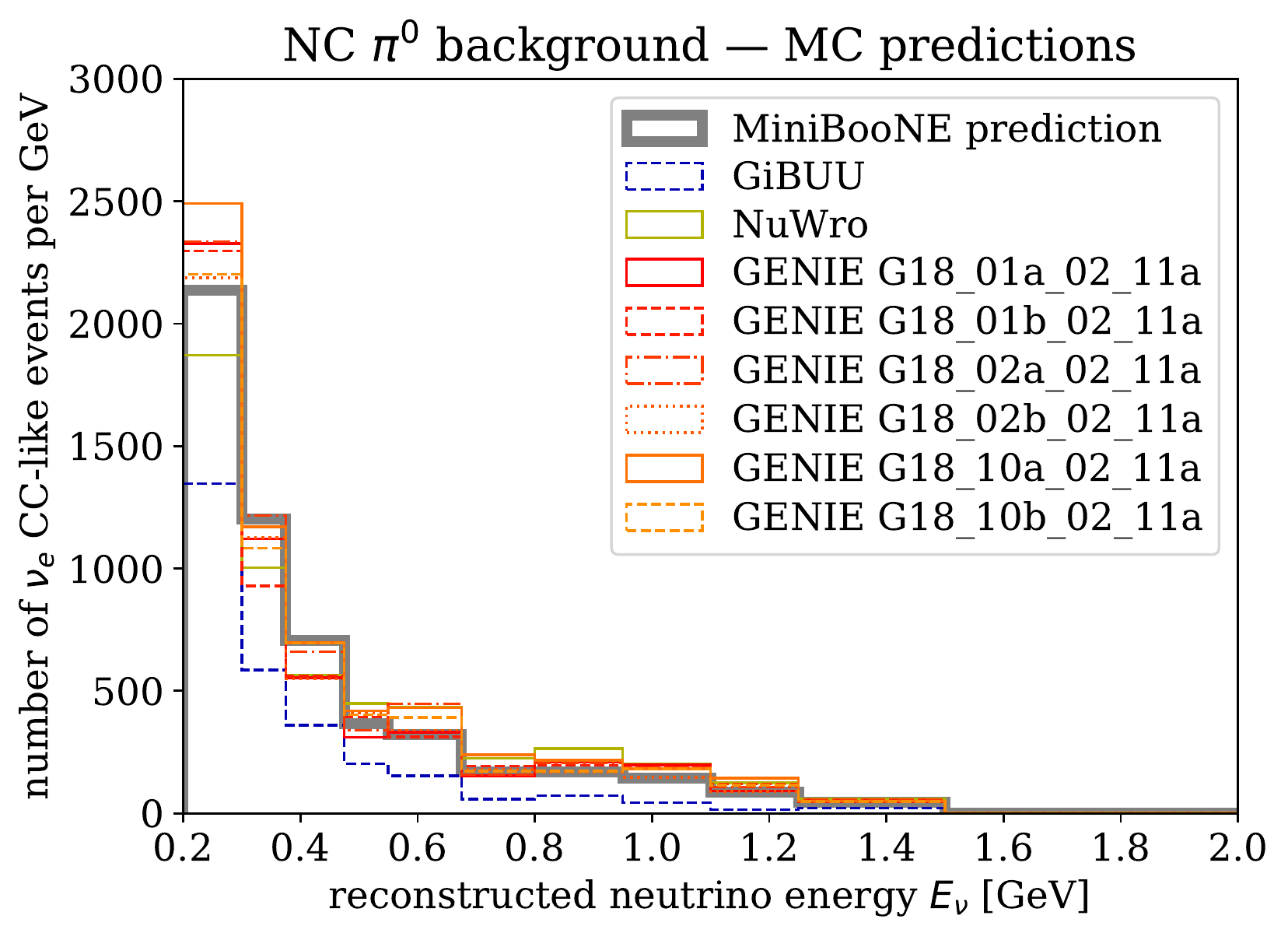} &
    \includegraphics[width=0.45\textwidth]{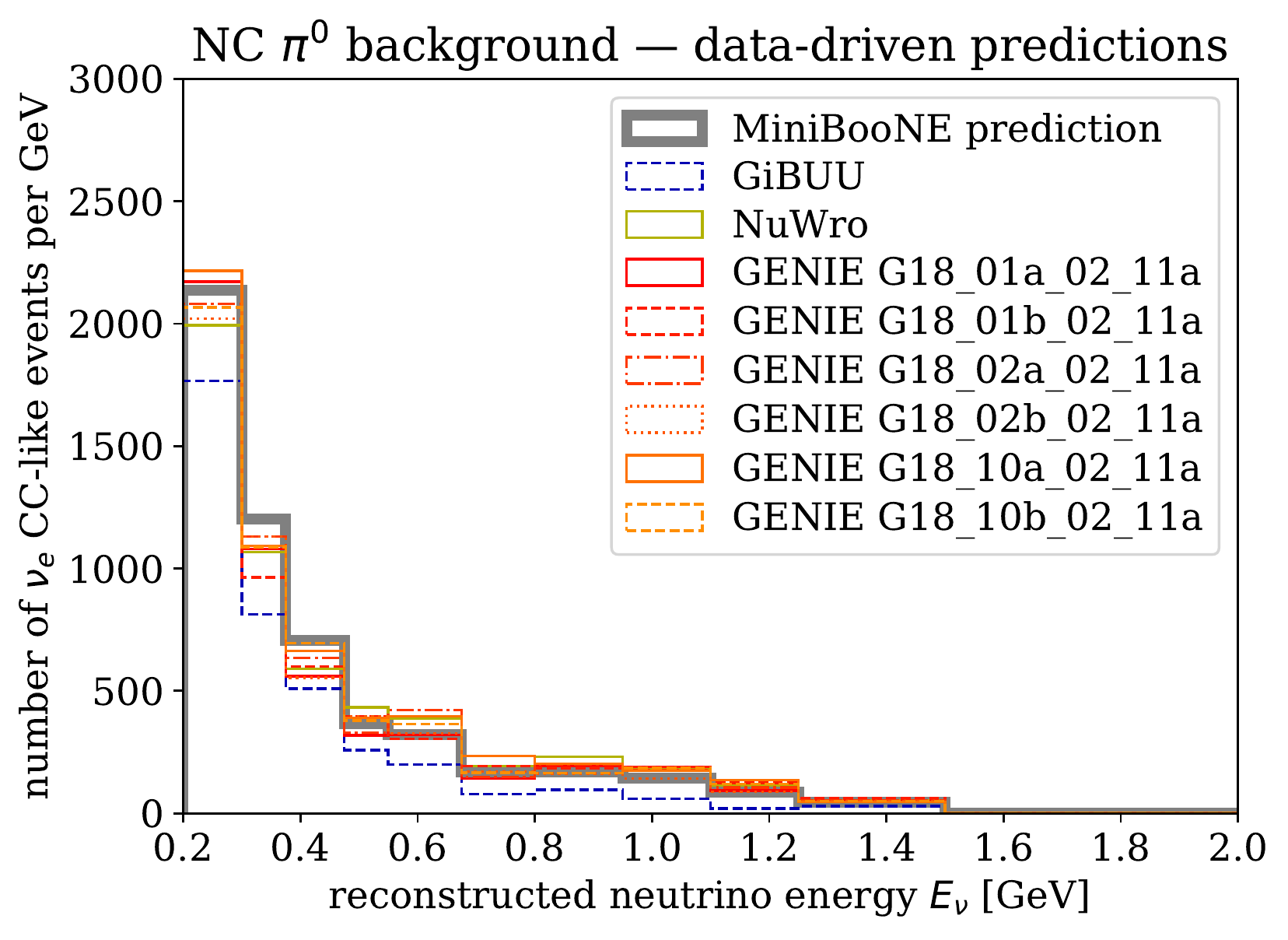} \\
  \end{tabular}
  \caption{Number of events for the $\pi^0$ channel as a function of reconstructed neutrino energy for several considered Monte Carlo event generators. The left panel shows the out-of-the-box event generator predictions while in the right panel data-driven predictions, where $\pi^0$ data is considered as well, are shown.}
  \label{fig:1}
\end{figure}

For $\mathcal{O}$(GeV) neutrino energies \cite{Formaggio:2012cpf}, neutrino-nucleus interactions still feature relatively large uncertainties in the cross sections. It is therefore worthwhile to investigate whether different nuclear models can alleviate or possibly even explain the anomaly by predicting more events in the aforementioned three channels. Such an analysis was performed in \cite{Brdar:2021ysi} where nuclear and hadronic physics uncertainties have been explored. Different models are implemented in several Monte Carlo event generators; among the latter, GENIE \cite{Andreopoulos:2009rq}, GiBUU \cite{Leitner:2008ue} and NuWro \cite{Golan:2012wx} are considered in \cite{Brdar:2021ysi}. The differences in the neutrino event rates across models can be inferred from Fig.~\ref{fig:1} for the $\pi^0$ channel.
The upshot of the analysis is that the MiniBooNE anomaly gets alleviated for some nuclear models to $\lesssim 4\sigma$; this, however, also implies that the explanation of the anomaly within the Standard Model is not feasible. Let us point out that in the recent work \cite{Kelly:2022uaa} another previously unconsidered process for the appearance of a single shower was studied and the authors also further explored the $\pi^0$ channel.

The results in \cite{Brdar:2021ysi,Kelly:2022uaa} motivate future and justify previous BSM considerations of the MiniBooNE anomaly. The most widely considered BSM scenario for the explanation of MiniBooNE anomaly is eV-scale sterile neutrino which 
poses as a ``catalyst'' for an efficient transition between muon and electron neutrinos where the latter induce electron-like events in the detector.
This scenario requires relatively large sterile neutrino mixing angles. While MiniBooNE anomaly can be explained in such a framework \cite{MiniBooNE:2018esg,MiniBooNE:2020pnu}, this explanation is disfavored once all available data is included \cite{Dentler:2018sju,Gariazzo:2017fdh,Hardin:2022muu}; this is mostly driven by MINOS and IceCube muon neutrino disappearance data \cite{MINOS:2017cae,IceCube:2020phf}. While the consistency of the eV-scale sterile neutrino explanation can be improved by adding more BSM ingredients \cite{Babu:2022non}, let us stress the existence of many non-oscillatory solutions; see \cite{Acero:2022wqg} for the summary of proposed models and \cite{Brdar:2020tle} for a model-independent approach.

 The MiniBooNE anomaly remains a puzzle which will hopefully not outlive the SBN \cite{MicroBooNE:2021zai,MicroBooNE:2021tya} with ICARUS \cite{Rubbia:2011ft}, MicroBooNE \cite{MicroBooNE:2016pwy} and SBND \cite{McConkey:2017dsv} experiments.


\section{Gallium Anomaly}
\label{sec:gallium}
In gallium experiments, radioactive source (typically $^{51}$Cr which decays via electron capture) produces a strong flux of electron neutrinos. The process of neutrino capture on $^{71}$Ga leads to the production of $^{71}$Ge which is then extracted using experimental techniques. The observed $\sim 20\%$ deficit of events corresponds to $\gtrsim 5\sigma$ anomaly, which emerged chiefly due to recent measurements from BEST \cite{Barinov:2021asz}. An obvious solution to such an observation is the model with eV-scale sterile neutrino to which electron neutrinos would partially oscillate. However, in order to explain a $20\%$ deficit, the mixing angle would need to be rather large \cite{Barinov:2022wfh} and that is comfortably disfavored by solar and reactor experiments \cite{Giunti:2021kab,Berryman:2021yan}. Hence, the gallium anomaly calls for an alternative explanation, either within or beyond the Standard Model.

\begin{figure}
  \centering
  \begin{tabular}{c@{\quad\quad}c}
    \includegraphics[width=0.40\textwidth]{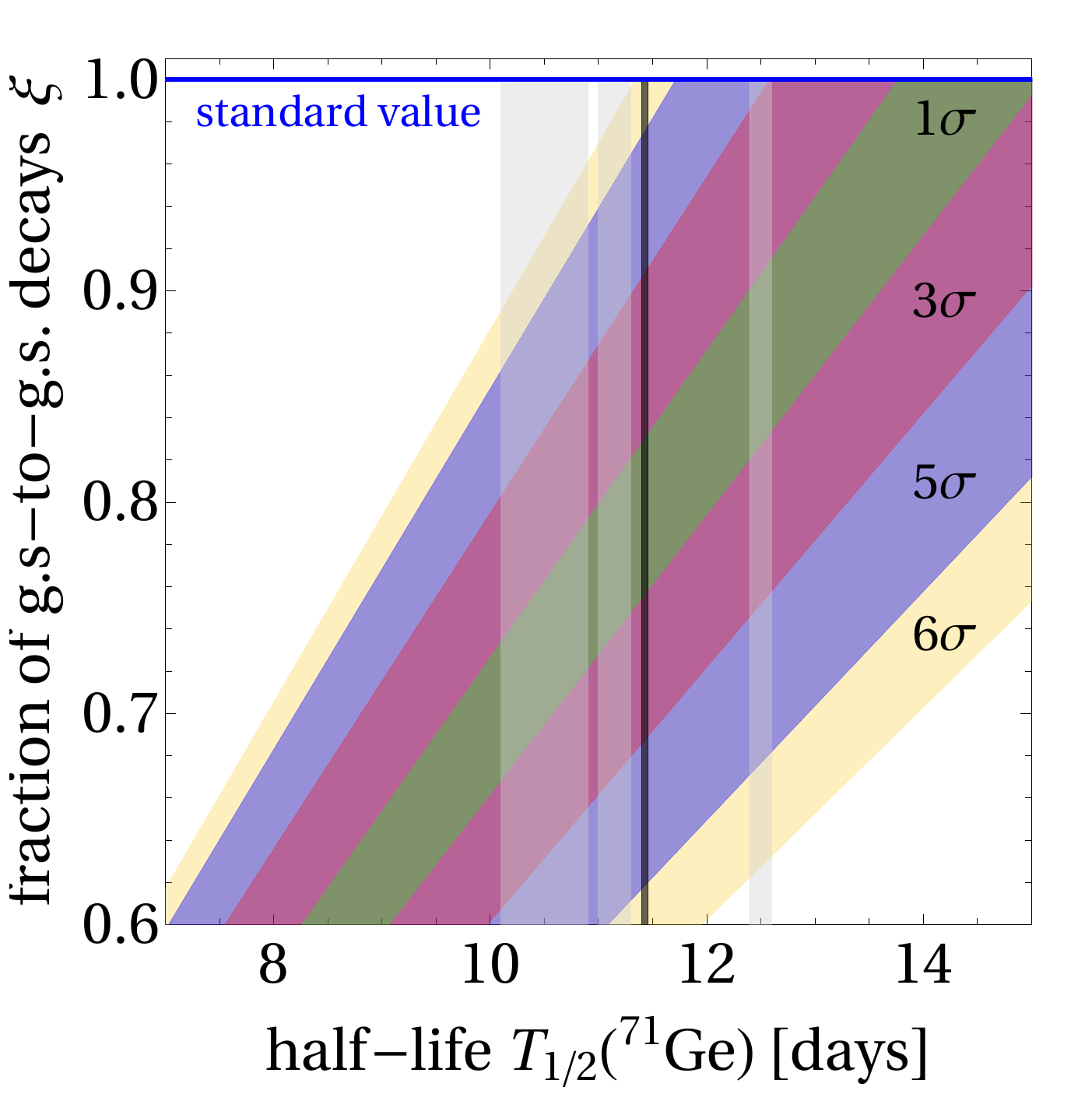} &
    \includegraphics[width=0.419\textwidth]{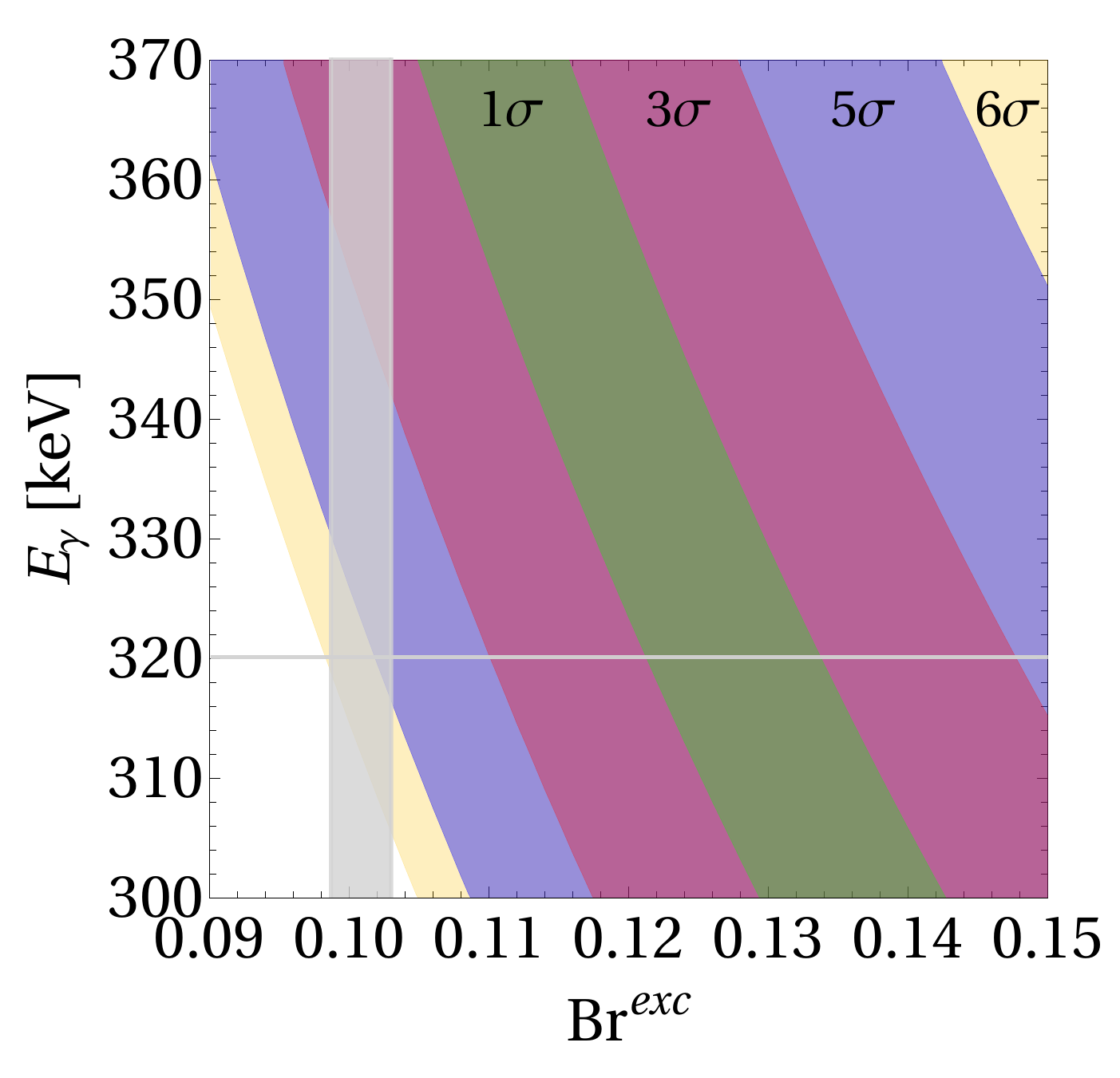} \\
  \end{tabular}
  \caption{{\it{Left}}: Statistical significance of the gallium anomaly as a function of the $^{71}$Ge half-life and the fraction of $^{71}$Ge decays into the ground state of $^{71}$Ga. {\it{Right}}: Statistical significance of the gallium anomaly as a function of the branching ratio for the decay into the excited state of $^{51}$V  and the energy of the emitted $\gamma$-ray. 
}
  \label{fig:2}
\end{figure}
 An explanation of the anomaly can be sought in the cross section for  $\nu_e + ^{71}$Ga $\to e^- + ^{71}$Ge. Given that $^{71}$Ge decays via electron capture, the matrix element is the same as for the $\nu_e$ capture on gallium. Therefore, the cross section of interest can be determined via measurement of $^{71}$Ge half-life. The most precise measurement to date reads $11.43\pm 0.03$ days \cite{Hampel:1985zz} and it is this half-life that is being employed when claiming $\sim 5\sigma$ deviation from the expected event rate. While the result from \cite{Hampel:1985zz} appears robust since it is obtained by performing several measurements with two different methods, one should still point out that there are three other measurements that found different results for the half-life (see gray vertical bands in the left panel of Fig.~\ref{fig:2}); in particular, if the one that corresponds to the largest half-life is taken at face value, the gallium anomaly would be alleviated to $\sim 3\sigma$ \footnote{The impact of $^{71}$Ge half-life to the gallium anomaly was studied also in \cite{Giunti:2022xat}.}. Another way on how to overestimate matrix element for neutrino capture on gallium is a  scenario in which $^{71}$Ge in 20\% of the cases decays into as yet undiscovered excited state(s) of $^{71}$Ga, existence of which is admittedly not supported by the nuclear data. The dependence of the statistical significance of the anomaly on such a scenario as well as on the $^{71}$Ge half-life is shown in the left panel of Fig.~\ref{fig:2}.
 
Another potential avenue for the explanation of the anomaly is the neutrino flux. The $^{51}$Cr source intensity is determined calorimetrically. $^{51}$Cr decays into $^{51}$V via electron capture and in roughly 10\% of those decays excited state of $^{51}$V is produced. This results in an emission of a 320 keV $\gamma$-ray. For the explanation of the anomaly, the branching ratio to the excited state should be roughly 2\% larger \cite{Brdar:2023cms}, see right panel of Fig.~\ref{fig:2} where the dependence of the significance of the anomaly on the energy of the emitted $\gamma$-ray is shown as well.  
Gallium anomaly can also be explained if extraction efficiency of $^{71}$Ge is $\sim 20\%$ smaller than claimed $\approx$ 95\% \cite{Barinov:2022wfh}; see detailed discussion in \cite{Brdar:2023cms}.

\begin{figure}
  \centering
  \includegraphics[width=0.55\textwidth]{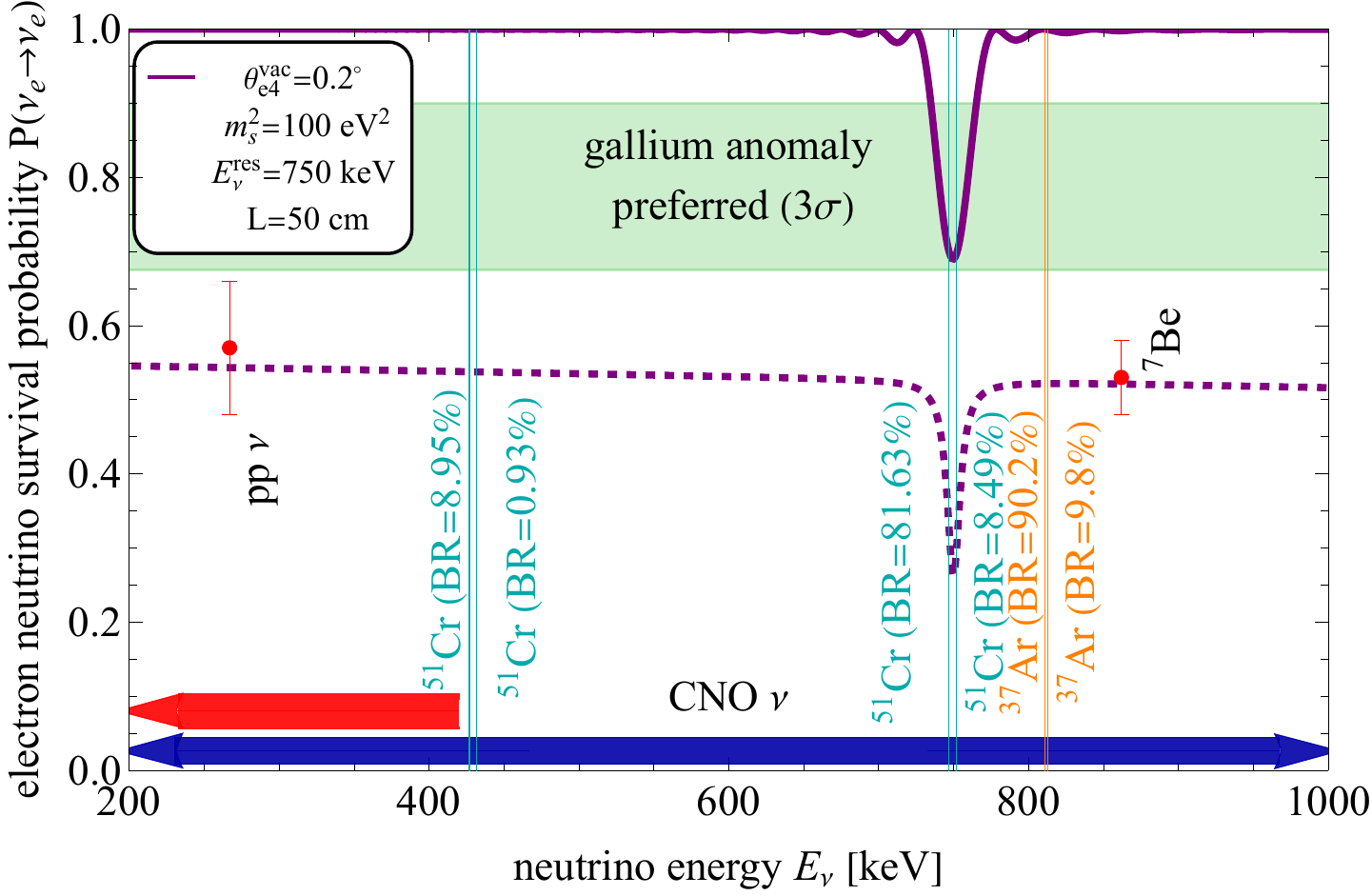}
  \caption{Survival probability of electron neutrino for the model in which MSW resonance is induced. The resonance appears in a very narrow energy window and hence is not constrained by solar neutrino data (see pp and $^{7}$Be data points).}
  \label{fig:3}
\end{figure}

Regarding BSM, as discussed above, vanilla eV-scale sterile neutrino is strongly disfavored. However, if sterile neutrino mixing angle can feature an enhancement at energies corresponding to those of neutrinos emitted from $^{51}$Cr, gallium anomaly could be explained without tension with solar and reactor data. This is 
possible by utilizing Mikheyev-Smirnov-Wolfenstein (MSW) resonance \cite{Wolfenstein:1977ue,Mikheyev:1985zog}. For this particular case, it was found that the resonant sterile neutrino mixing angle enhancement is achieved by introducing sterile neutrino interaction with ultralight dark matter or dark energy \cite{Brdar:2023cms}. This is illustrated in Fig.~\ref{fig:3} where the electron neutrino survival probability is shown. $^{51}$Cr emits neutrinos at four discrete energies; the most intense emission line is around 750 keV and, as seen from the figure, that is the energy where the resonance was achieved. The constraints form solar experiments are evaded due to a very narrow resonance; note, however, that forthcoming precise measurements of CNO neutrinos can probe this scenario. In order to evade limits from cosmology, it was found that sterile neutrino should decay \cite{Brdar:2023cms}. Further details about this model, as well as several other options for explaining the anomaly with BSM physics (e.g. via parametric resonance \cite{{Losada:2022uvr}}) may be found in \cite{Brdar:2023cms}.


\section{Summary and Conclusions}
Several short baseline anomalies still remain unsolved and, among those, MiniBooNE and gallium anomalies stand out as statistically the  most significant ones. Recently, it was demonstrated that nuclear and hadronic physics uncertainties can mildly alleviate but not fully resolve the MiniBooNE anomaly. The MiniBooNE anomaly is currently being tested at the SBN, using three detectors placed at different distances from the neutrino source. This experimental program is expected to have the final word on both oscillatory and non-oscillatory BSM explanations of the MiniBooNE anomaly.

 The gallium anomaly is another mystery whose explanation was recently scrutinized both from the Standard Model and the new physics perspective. Essentially, all proposed explanations can be tested by performing measurements with a different neutrino source (e.g.  $^{37}$Ar or $^{65}$Zn) and/or another detection material (e.g. $^{37}$Cl). Specifically, in the BSM scenario with the tuned MSW resonance, no deficit in the event rate is expected for the measurement with a $^{65}$Zn source that is being actively considered by BEST collaboration.   

\end{document}